\documentstyle[preprint,aps]{revtex}
\input epsf.tex
\def\DESepsf(#1 width #2){\epsfxsize=#2 \epsfbox{#1}}
\begin{document}
\preprint{\vbox{\hbox{OITS-566}}}
\draft
\title{ CP asymmetry Relations Between\\
  $\bar B^0\rightarrow \pi\pi$ And
$\bar B^0\rightarrow \pi K$ Rates
\footnote{Work supported in part by the Department of Energy Grant No.
DE-FG06-85ER40224.}}
\author{N.G. Deshpande, and Xiao-Gang He}
\address{Institute of Theoretical Science\\
University of Oregon\\
Eugene, OR 97403-5203, USA}

\date{December 1994}
\maketitle
\begin{abstract}
We prove that CP violating rate difference $\Delta (\bar B^0 \rightarrow
\pi^+\pi^-)
= \Gamma (\bar B^0 \rightarrow \pi^+\pi^-)
- \Gamma ( B^0 \rightarrow \pi^-\pi^+)$ is related to
 $\Delta (\bar B^0 \rightarrow \pi^+ K^-) = \Gamma (\bar B^0 \rightarrow
\pi^+K^-)
- \Gamma (B^0 \rightarrow \pi^-K^+)$ in the three generation Standard Model.
Neglecting small annihilation diagrams, and in the SU(3)
symmetry limit,  we show $\Delta (\bar B^0 \rightarrow \pi^+\pi^-)
= - \Delta (\bar B^0 \rightarrow \pi^+ K^-)$. The SU(3) breaking effects are
estimated using factorization approximation, and yield $\Delta (\bar
B^0\rightarrow \pi^+\pi^-) \approx -(f_\pi/f_K)^2\Delta (\bar B^0 \rightarrow
\pi^+K^-)$. Usefulness of this remarkable relation for determining phases in
the CKM unitarity triangle is discussed.

\end{abstract}
Detection of CP violation and verification of the unitarity triangle of the CKM
matrix is a major goal of $B$ factories. Measurement of rate asymmetry in
certain channels not only establishes direct CP violation, but can aid in
determining some of the angles of the unitarity triangle. In this letter we
shall prove a remakable relationship between rate difference
 $\Delta (\bar B^0 \rightarrow \pi^+\pi^-(\pi^0\pi^0))
= \Gamma (\bar B^0 \rightarrow \pi^+\pi^-(\pi^0\pi^0))
- \Gamma ( B^0 \rightarrow \pi^-\pi^+(\pi^0\pi^0))$ and
 $\Delta (\bar B^0 \rightarrow \pi^+ K^- (\pi^0\bar K^0)) = \Gamma (\bar B^0
\rightarrow \pi^+K^- (\pi^0\bar K^0))
- \Gamma (B^0 \rightarrow \pi^-K^+(\pi^0 K^0))$. This relationship follows
purely from SU(3) symmetry and ignoring annihilation diagrams which
can be shown to make negligible contributions. The usefulness of such a
relationship
lies in that difficult to measure rate difference like $\Delta(\bar
B^0\rightarrow \pi^+\pi^-)$ can be related to easier measurement of
$\Delta(\bar B^0 \rightarrow \pi^+ K^-)$ which is a self-tagging mode.
Similarly, it might prove easier to measure $\Delta (\bar B^0 \rightarrow
\pi^0\bar K^0)$ than the rate difference of the suppressed mode $\bar B^0
\rightarrow \pi^0\pi^0$.

In the Standard Model (SM) the effective Hamiltonian for
$B \rightarrow\pi\pi$ and $B\rightarrow\pi K$ decays
can be written as follows:
\begin{eqnarray}
H_{eff}^q &=& {G_F\over \sqrt{2}}[V_{ub}V^*_{uq}(c_1O_1^q + c_2 O_2^q) -
\sum_{i=3}^{10}(V_{ub}V^*_{uq} c_i^u\nonumber\\
&+&V_{cb}V^*_{cq} c_i^c
+V_{tb}V^*_{tq} c_i^t)
O_i^q] +H.C.\;,
\end{eqnarray}
where the Wilson coefficients $c_i^f$ are defined at the scale of $\mu \approx
m_b$ which have been evaluated to the next-to-leading order in QCD\cite{5}, the
superscript f indicates the loop contribuiton from f quark, and $O_i^q$ are
defined as
\begin{eqnarray}
O_1^q &=& \bar q_\alpha \gamma_\mu(1-\gamma_5)u_\beta\bar
u_\beta\gamma^\mu(1-\gamma_5)b_\alpha\;,\;O_2^q =\bar q
\gamma_\mu(1-\gamma_5)u\bar
u\gamma^\mu(1-\gamma_5)b\;,\nonumber\\
O_3^q &=&\bar q \gamma_\mu(1-\gamma_5)b
\bar q' \gamma_\mu(1-\gamma_5) q'\;,\;\;\;\;\;\;\;O_4^q = \bar q_\alpha
\gamma_\mu(1-\gamma_5)b_\beta
\bar q'_\beta \gamma_\mu(1-\gamma_5) q'_\alpha\;,\\
O_5^q &=&\bar q \gamma_\mu(1-\gamma_5)b  \bar q'
\gamma^\mu(1+\gamma_5)q'\;,\;\;\;\;\;\;\;O_6^q = \bar q_\alpha
\gamma_\mu(1-\gamma_5)b_\beta
\bar q'_\beta \gamma_\mu(1+\gamma_5) q'_\alpha\;,\nonumber\\
O_7^q &=& {3\over 2}\bar q \gamma_\mu(1-\gamma_5)b  e_{q'}\bar q'
\gamma^\mu(1+\gamma_5)q'\;,\;O_8^q = {3\over 2}\bar q_\alpha
\gamma_\mu(1-\gamma_5)b_\beta
e_{q'}\bar q'_\beta \gamma_\mu(1+\gamma_5) q'_\alpha\;,\nonumber\\
O_9^q &=& {3\over 2}\bar q \gamma_\mu(1-\gamma_5)b  e_{q'}\bar q'
\gamma^\mu(1-\gamma_5)q'\;,\;O_{10}^q = {3\over 2}\bar q_\alpha
\gamma_\mu(1-\gamma_5)b_\beta
e_{q'}\bar q'_\beta \gamma_\mu(1-\gamma_5) q'_\alpha\;.\nonumber
\end{eqnarray}
Here $q'$ is summed over u, d, and s. For $\Delta S = 0$ processes,
$q = d$, and for $\Delta S = 1$ processes, $q =s$. $O_{2}$, $O_1$ are the tree
level and QCD corrected operators. $O_{3-6}$ are the strong gluon induced
penguin operators, and operators $O_{7-10}$ are due to $\gamma$ and Z exchange,
and ``box'' diagrams at loop level.

Using the unitarity property of the CKM matrix, we can eliminate the term
proportional to $V_{cb}V^*_{cq}$ in the effective Hamiltonian. The $B$
decay amplitude due to the complex
effective Hamiltonian displayed above can be paramerized, without  loss of
generality, as
\begin{eqnarray}
<final\;state|H_{eff}^q|B> = V_{ub}V^*_{uq} T_q + V_{tb}V^*_{tq}P_q\;,
\end{eqnarray}
where $T_q$ contains the $tree\; contributions$ and $penguin\; contributions$
due to
u and c internal quarks, while $P_q$ only contains $penguin\; contribuitons$
from internal c and t quarks.

Since the effective Hamiltonian $H_{eff}^d$ responsible for $\Delta S = 0$ $B$
decays is related to $H^s_{eff}$ for $\Delta S = 1$ $B$ decays by just changing
d quark  to s quark, one expects certain relations
between $T_d$, $P_d$ and $T_s$, $P_s$ in the SU(3) limit. Let us consider the
two pseudoscalar meson decays of $B$ mesons.

The operators $Q_{1,2}$, $O_{3-6}$, and $O_{7-10}$ transform under SU(3)
symmetry as $\bar 3_a + \bar 3_b +6 + \bar {15}$,
$\bar 3$, and $\bar 3_a + \bar 3_b +6 + \bar {15}$, respectively. In general,
we can
write the SU(3) invaraint amplitude for $B$ to two octet pseudoscalar mesons
for  $T_q$ in the following form\cite{6}
\begin{eqnarray}
T&=& A_{(\bar 3)}^TB_i H(\bar 3)^i (M_l^k M_k^l) + C^T_{(\bar 3)}
B_i M^i_kM^k_jH(\bar 3)^j \nonumber\\
&+& A^T_{(6)}B_i H(6)^{ij}_k M^l_jM^k_l + C^T_{(6)}B_iM^i_jH(6
)^{jk}_lM^l_k\nonumber\\
&+&A^T_{(\bar {15})}B_i H(\bar {15})^{ij}_k M^l_jM^k_l + C^T_{(\bar
{15})}B_iM^i_j
H(\bar {15} )^{jk}_lM^l_k\;,
\end{eqnarray}
where $B_i = (B^-, B^0, B^0_s)$ is a SU(3) triplet, $M_{ij}$ is the SU(3)
pseudoscalar octet, and the
matrices H represent the transformation properties of the operators $O_{1
-10}$.
$H(6)$ is a traceless tensor that is antisymmetric on its upper indices, and
$H(\bar {15} )$ is also a traceless tensor but is symmetric on its
upper indices. For $q=d$, the non-zero entries of the H matrices are given by
\begin{eqnarray}
H(\bar 3)^2 &=& 1\;,\;\;
H(6)^{12}_1 = H(6)^{23}_3 = 1\;,\;\;H(6)^{21}_1 = H(6)^{32}_3 =
-1\;,\nonumber\\
H(\bar {15} )^{12}_1 &=& H(\bar {15} )^{21}_1 = 3\;,\; H(\bar {15} )^{22}_2 =
-2\;,\;
H(\bar {15} )^{32}_3 = H(\bar {15} )^{23}_3 = -1\;.
\end{eqnarray}
For $q = s$, the non-zero entries are
\begin{eqnarray}
H(\bar 3)^3 &=& 1\;,\;\;
H(6)^{13}_1 = H(6)^{32}_2 = 1\;,\;\;H(6)^{31}_1 = H(6)^{23}_2 =
-1\;,\nonumber\\
H(\bar {15} )^{13}_1 &=& H(\bar {15} ) ^{31}_1 = 3\;,\; H(\bar {15} )^{33}_3 =
-2\;,\;
H(\bar {15} )^{32}_2 = H(\bar {15} )^{23}_2 = -1\;.
\end{eqnarray}
We obtain the amplitudes $T_d(\pi\pi)$, $T_s(\pi K)$ for
$\bar B^0 \rightarrow \pi \pi$, $\bar B^0 \rightarrow \pi K$ as
\begin{eqnarray}
T_d(\pi^+\pi^-) &=& 2A^T_{(\bar 3)} +C^T_{(\bar 3)}
-A^T_{(6)} + C^T_{(6)} + A^T_{(\bar {15} )} + 3 C^T_{(\bar {15}
)}\;,\nonumber\\
T_d(\pi^0\pi^0) &=& {1\over \sqrt{2}} (2A^T_{(\bar 3)} +C^T_{(\bar 3)}
-A^T_{(6)} + C^T_{(6)} + A^T_{(\bar {15} )} -5 C^T_{(\bar {15}
)})\;,\nonumber\\
T_d(\pi^-\pi^0) &=& {8\over \sqrt{2}}C^T_{(\bar {15} )}\;,\nonumber\\
T_s(\pi^+ K^-) &=& C^T_{(\bar 3)}
-A^T_{(6)} + C^T_{(6)} - A^T_{(\bar {15} )} + 3 C^T_{(\bar {15}
)}\;,\nonumber\\
T_s(\pi^0\bar K^0) &=& -{1\over \sqrt{2}} (C^T_{(\bar 3)}
-A^T_{(6)} + C^T_{(6)} - A^T_{(\bar {15} )} -5 C^T_{(\bar {15} )})\;.
\end{eqnarray}
We also have similar relations for the amplitude $P_q$. The corresponding
amplitude will be denoted
by $A^P_i$ and $C^P_i$.
We clearly see the triangle relation (which follows from isospin) holds:
\begin{eqnarray}
A(\bar B^0\rightarrow \pi^0\pi^0) + A(B^-\rightarrow \pi^-\pi^0) &=& {1\over
\sqrt{2}} A(\bar B^0\rightarrow \pi^+\pi^-)\;.
\end{eqnarray}
As also a similar relation for the charge conjugate decay modes.

The amplitudes $A_{(\bar 3), (6), (\bar {15} )}$ all correspond to
annihilation contributions. This can be verified because the light quark
index in the $B$ meson is contracted with the Hamiltonian.
In the factorization approximation, these amplitudes correspond to the matrix
element of the form, for example for $\bar B^0 \rightarrow \pi^+\pi^-$ decay,
\begin{eqnarray}
M=<0|\bar d \Gamma^1 b|\bar B^0> <\pi^+\pi^-|\bar q \Gamma_2 q|0>\;,
\end{eqnarray}
where $q$ can be u or d quarks.
If $\Gamma^1 = \gamma_\mu (1-\gamma_5)$, and $\Gamma_2 = \gamma^\mu (1\pm
\gamma_5)$, this matrix element is equal to zero due to vector current
conservation. The only exception is when the operators are Fierz transformed,
one also obtains a contribution of the type, $\Gamma^1 = 1-\gamma_5$, and
$\Gamma_2 = 1+\gamma_5$. However, this contribution  is suppressed compared
with other contributions.
In the factorization approximation, for $q = d$, this contribution is given by,
\begin{eqnarray}
M = if_B m_B^2{m_\pi^2\over m_u+m_d}{1\over m_b+m_d} F^{\pi\pi}(m_B^2)\;,
\end{eqnarray}
where we have used, $<0|\bar d (1-\gamma_5) b|\bar B^0> = if_B m_B^2/(m_b+m_d)$
and $<\pi^+\pi^-|\bar d d|0> = F^{\pi\pi}(q^2)m_\pi^2/(m_u+m_d)$\cite{7}.
Assuming single pole model for the form factor, $F^{\pi\pi}(q^2) = 1/(1-
q^2/m_\sigma^2)$
with $m_\sigma = 700$ MeV, $F^{\pi\pi}(m_B^2) \approx -0.02$.
For $\bar B^0\rightarrow \pi^+\pi^-$ we find that the
annihilation contribution  to $P_d(\pi^+\pi^-)$ is only about 4\%, and the
contribution to $T_d(\pi^+\pi^-)$ is much smaller.
To a good approximation all annihilation amplitudes $A_{(\bar 3),(6), (\bar
{15})}$ can be neglected. From now on we will work in this approximation. We
obtain:
\begin{eqnarray}
T_{+-} = T_d(\pi^+\pi^-) = T_s(\pi^+ K^-)\;,\;\;\; P_{+-} = P_d(\pi^+\pi^-) =
P_s(\pi^+K^-)\;,\nonumber\\
T_{00} = T_d(\pi^0\pi^0) = -T_s(\pi^0\bar K^0)\;,\;\;\; P_{00} =
P_d(\pi^0\pi^0) = - P_s(\pi^0 \bar K^0)\;,
\end{eqnarray}
and
\begin{eqnarray}
A(\bar B^0 \rightarrow \pi^+\pi^-) &=& V_{ub}V_{ud}^* T_{+-} + V_{tb}V^*_{td}
P_{+-}\;,\nonumber\\
A(\bar B^0 \rightarrow \pi^+K^-) &=& V_{ub}V_{us}^* T_{+-} + V_{tb}V^*_{ts}
P_{+-}\;,\nonumber\\
A(\bar B^0 \rightarrow \pi^0\pi^0) &=& V_{ub}V_{ud}^* T_{00} + V_{tb}V^*_{td}
P_{00}\;,\nonumber\\
A(\bar B^0 \rightarrow \pi^0\bar K^0) &=& -V_{ub}V_{us}^* T_{00} -
V_{tb}V^*_{ts} P_{00}\;.
\end{eqnarray}

Analogus relations have been discussed in the context of obtaining information
about penguin contributions to $B$ decays
 and to determine the unitarity
triangle of the CKM matrix\cite{8}. These studies suffer from uncertainties in
the strong rescattering phases in the amplitudes. Our derivation spell out
the precise assumptions that are necessary to obtain the relations. We shall
use them to derive relations between the decay rate differences which do not
have uncertainties associated with lack of knowledge of the strong rescattering
phases. We have
\begin{eqnarray}
\Delta(\bar B^0\rightarrow \pi^+\pi^-) &=&  -4Im(V_{ub}V^*_{ud}V_{tb}^*V_{td})
Im(T_{+-}P^*_{+-}){m_B\lambda_{\pi\pi}\over 16 \pi}\;,\nonumber\\
\Delta(\bar B^0\rightarrow \pi^+K^-)&=& -4Im(V_{ub}V^*_{us}V_{tb}^*V_{ts})
Im(T_{+-}P^*_{+-}){m_B\lambda_{\pi K}\over 16 \pi}\;,\nonumber\\
\Delta(\bar B^0\rightarrow \pi^0\pi^0) &=&  -4Im(V_{ub}V^*_{ud}V_{tb}^*V_{td})
Im(T_{00}P^*_{00}){m_B\lambda_{\pi\pi}\over 16 \pi}\;,\nonumber\\
\Delta(\bar B^0\rightarrow \pi^0\bar K^0)&=& -4Im(V_{ub}V^*_{us}V_{tb}^*V_{ts})
Im(T_{00}P^*_{00}){m_B\lambda_{\pi K}\over 16 \pi}\;,\nonumber\\
\end{eqnarray}
where  $\lambda_{ab} = \sqrt{1-2(m_a^2+m_b^2)/m_B^2 + (m_a^2-m_b^2)^2/m_B^4}$.
In the SU(3) symmetry limit, $\lambda _{\pi\pi} = \lambda_{\pi K}$.
Due to the unitarity property of the CKM matrix, for three generations
of quarks, $Im(V_{ub}V^*_{us}V_{tb}^*V_{ts}) =
-Im(V_{ub}V^*_{ud}V_{tb}^*V_{td})$. We then find
\begin{eqnarray}
\Delta(\bar B^0 \rightarrow \pi^+\pi^-)&=& -\Delta(\bar B^0\rightarrow
\pi^+K^-)\;,\nonumber\\
\Delta(\bar B^0 \rightarrow \pi^0\pi^0)&=& -\Delta(\bar B^0\rightarrow \pi^0
\bar K^0).
\end{eqnarray}
These non-trivial equality relations do not
depend on the numerical values of the final state rescattering phases.
Of course these relations are true only for three
generation model. Therefore they also provide  tests for the three generation
model.

In the real world the SU(3) symmetry is not exact. The relations obtained above
will be modified. We estimate the SU(3) symmetry breaking effects by specific
calculations in the factorization approximation. In this approxmation,
we have
\begin{eqnarray}
T_d(\pi^-\pi^+) &=& i{G_F\over \sqrt{2}}f_{\pi}F^{B\pi}_0(m_\pi^2)
(m_B^2-m_\pi^2)[\xi c_1 +c_2
+\xi c_3^{cu} +c_4^{cu} + \xi c_9^{cu}
+c_{10}^{cu}\nonumber\\
&+& {2m_\pi^2 \over (m_b-m_u)(m_u+m_d)}
(\xi c_5^{cu} +c_6^{cu} +\xi c_7^{cu} + c_8^{cu})]\;,\nonumber\\
T_s(\pi^+ K^-) &=& i{G_F\over \sqrt{2}}f_{K}F^{B\pi}_0(m_K^2)
(m_B^2-m_\pi^2)
[\xi c_1 +c_2 +\xi c_3^{cu} +c_4^{cu} +\xi c_9^{cu}
+c_{10}^{cu}\nonumber\\
& +& {2 m_K^2 \over (m_b-m_u)(m_u+m_s)}
(\xi c_5^{cu} +c_6^{cu} +\xi c_7^{cu} + c_8^{cu})]\;,\nonumber\\
T_d(\pi^0\pi^0) &=& i{G_F\over \sqrt{2}}f_{\pi}F^{B\pi}_0(m_\pi^2)
(m_B^2-m_\pi^2)[-c_1 -\xi c_2
+\xi c_3^{cu} +c_4^{cu}\nonumber\\
& +& {3\over 2}(c_7^{cu} +\xi c_8^{cu} - c_9^{cu}
-\xi c_{10}^{cu}) -{1\over 2}(\xi c_9^{cu}
+c_{10}^{cu})\nonumber\\
&+& {2m_\pi^2 \over (m_b-m_d)(2m_d)}
(\xi c_5^{cu} +c_6^{cu} -{1\over 2}(\xi c_7^{cu} + c_8^{cu}))]\;,\nonumber\\
T_s(\pi^0 \bar K^0) &=& i{G_F\over \sqrt{2}}\{f_{\pi}F^{BK}_0(m_{\pi}^2)
(m_B^2-m_K^2)
[ c_1 + \xi c_2 -{3\over 2}(c_7^{cu} +\xi c_8^{cu} - c_9^{cu}
-\xi c_{10}^{cu}) ]\nonumber\\
&-&if_K F^{B\pi}_0(m_K^2)(m_B^2-m_\pi^2)[c_3^{cu}+\xi c_4^{cu} - {1\over 2}
(\xi c_9^{cu}+c_{10}^{cu})\nonumber\\
& -& {2 m_K^2 \over (m_b-m_d)(m_d+m_s)}
(\xi c_5^{cu} +c_6^{cu} -{1\over 2}(\xi c_7^{cu} + c_8^{cu}))]\}\;,
\end{eqnarray}
where $c_i^{cu} = c_i^c-c_i^u$, $c_i^{ct} = c_i^c-c_i^t$,  and $\xi = 1/N_c$
with $N_c$ being the number of color. The amplitude
$P_{d,s}$ are obtained by setting $c_{1,2} = 0$ and changing
$c_i^{cu}$ to $c_i^{ct}$.  We have used the following decompositions for the
form factors
\begin{eqnarray}
&<&\pi^+(q)|\bar d \gamma_\mu(1-\gamma_5) u|0> = if_\pi q_\mu\;,
<K^+(q)|\bar d \gamma_\mu(1-\gamma_5) u|0> = if_K q_\mu\;,\nonumber\\
&<&\pi^-(k)|\bar u \gamma_\mu b|\bar B^0(p)> = (k+p)_\mu
F^{B\pi}_1+(m_\pi^2-m_B^2){q_\mu\over q^2}(F^{B\pi}_1(q^2)
-F^{B\pi}_0(q^2))\;,\nonumber\\
&<&K^-(k)|\bar u \gamma_\mu b|\bar B^0(p)> = (k+p)_\mu
F^{BK}_1+(m_\pi^2-m_B^2){q_\mu\over q^2}(F^{BK}_1(q^2)
-F^{BK}_0(q^2))\;.
\end{eqnarray}
In the above we have neglected all annihilation contribuitons which are small
compared with other contribuitons as discussed earlier.

Using the fact $m_\pi^2/(m_u+m_d) = m_K^2/(m_u+m_s)$, we obtain
\begin{eqnarray}
\Delta(\bar B^0\rightarrow \pi^+\pi^-) =
-{(f_\pi F^{B\pi}_0(m_\pi^2))^2\over
(f_{K} F^{B\pi}_0(m_K^2))^2}{\lambda_{\pi\pi}\over \lambda_{\pi K}}\Delta(\bar
B^0\rightarrow \pi^+K^-),
\end{eqnarray}
It is clear that in the SU(3) symmetry limit, the above relation reduces to
eq.(14).
Now we need to use the physical masses for $\pi$ and $K$. Assuming single
pole for the form factor $F^{B\pi}_0(q^2)$, the form factor has the form
$F^{B\pi}_0(q^2) = 1/(1 - q^2/m^2_{0^+})$ with $m_{0^+} = 5.78$ GeV.
To a good approximation, we have $(\lambda_{\pi\pi}/\lambda_{\pi
K})(F_0^{B\pi}(m_\pi^2)/F_0^{B\pi}(m_K^2))^2 \approx 1$.  We finally obtain
\begin{eqnarray}
\Delta(\bar B^0\rightarrow\pi^+\pi^-) \approx - {f_\pi^2\over f_K^2}\Delta(\bar
B^0\rightarrow \pi^+K^-).
\end{eqnarray}

For $\bar B^0 \rightarrow \pi^0\pi^0$ and $\bar B^0 \rightarrow \pi^0 \bar
K^0$,
the correction is more complicated for two reasons: i) in general $f_\pi
F^{BK}_0(m_\pi^2)$ is not equal to $f_K F^{B\pi}_0(m_K^2)$, and ii) the u and d
quark masses are not equal. These cause the amplitudes $T(P)_d(\pi^0\pi^0)$
and $T(P)_s(\pi^0\bar K^0)$ for
$\bar B^0 \rightarrow \pi^0\pi^0$ and $\bar B^0\rightarrow \pi^0 \bar K^0$ to
be different not simply by an overall factor as in the case for
$\bar B^0 \rightarrow \pi^+\pi^-$ and $\bar B^0 \rightarrow \pi^+ K^-$. However
we estimate that the SU(3) breaking effect is about 30\%.

The relations obtained above will provide useful means of measuring a phase
angle in the unitarity triangle of the CKM matrix.
The angle $\alpha = arc(V_{tb}V_{td}^*/V_{ub}V_{ud}^*)$, can be determined by
measuring the time dependent CP asymmetry $a(t)_{+-(00)}$ in
$\bar B^0(B^0) \rightarrow \pi^+\pi^-(\pi^0\pi^0)$ decays\cite{1,2}.
The coefficient of the term varying with time as $\mbox{sin} (\Delta m t)$ is
proportional to $Im \lambda_{+-(00)}$ where
\begin{eqnarray}
\lambda_{+-(00)} = (V_{tb}^*V_{td}/V_{tb}V_{td}^*) {A( \bar B^0\rightarrow
\pi^+\pi^-(\pi^0\pi^0))\over
A( B^0\rightarrow \pi^-\pi^+(\pi^0\pi^0))}\;.
\end{eqnarray}
If penguin contributions are ignored,
one finds $A(\bar B^0\rightarrow \pi^+\pi^-)/
A( B^0\rightarrow \pi^-\pi^+) = V_{ub}V_{ud}^*/V_{ub}^*V_{ud}$, and
$\mbox{Im}\lambda_{+-(00)} = -\mbox{sin}(2\alpha)$.
However, the penguin contributions have been shown to be important\cite{3} and
can not be ignored. The relation changes to:
\begin{eqnarray}
\mbox{Im}\lambda_{+-(00)} = -{|A( B^0\rightarrow \pi^+\pi^-)|\over
|A(\bar B^0\rightarrow \pi^-\pi^+)|} \mbox{sin}(2\alpha+\theta_{+-(00)})\;.
\end{eqnarray}
A method has been suggested to remove uncertainties due to this change by
depermining
$\theta_{+-(00)}$ which involves
reconstruction of the triangle relation of eq.(8)\cite{2}. This requires
precise
measurement of rate difference
 $\Delta( \bar B^0\rightarrow \pi^+\pi^-(\pi^0\pi^0))$.
It is difficult to measure these rate differences because all the decay modes
require tagging.
The rate difference $\Delta(\bar B^0\rightarrow \pi^+K^-)$,
 on the other hand, is much easier to measure.
Similarly, we can get information for $\Delta( \bar B^0\rightarrow \pi^0\pi^0)$
from the measurement of $\Delta( \bar B^0\rightarrow \pi^0\bar K^0)$.
In this case the rate difference $\Delta (\bar B^0\rightarrow \pi^0\bar K^0)$
is also a
difficult quantity to measure because it also needs tagging. However
it might be easier to measure compared with $\Delta(\bar B^0\rightarrow
\pi^0\pi^0)$ since $\bar B^0\rightarrow \pi^0\pi^0$ is expected to be
highly suppressed.

We would also like to remark that  if in the future $\Delta(\bar B^0\rightarrow
\pi^+ K^- (\pi^0 \bar K^0))$ and  $\Delta(\bar B^0\rightarrow
\pi^+\pi^-(\pi^0\pi^0))$ are all measured precisely, the relations obtained
above
will provide tests for the three generation model because additional
contributions to the decay rates from physics beyond the
three generation SM will
change these relations.


\begin{references}
\bibitem{5} N. G. Deshpande and Xiao-Gang He, Phys. Lett. {\bf B336},
(1994)471.
 R. Fleischer, Z. Phys. {\bf C62}, 81(1994); A. Buras, M. Jamin, M.
Lautenbacher and P. Weisz, Nucl. Phys.
{\bf B400}, 37(1993); A. Buras, M. Jamin and M. Lautenbacher, ibid, 75(1993);
M. Ciuchini, E. Franco, G. Martinelli and L. Reina, Nucl. Phys. {\bf B415},
403(1994).
\bibitem{6} M. Savage and M. Wise, Phys. Rev. {\bf D39}, (1989)3346.
\bibitem{7} A.I. Vainshtein, V.I. Zakharov and M.A. Shifman, Sov. Phys. JETP
{\bf 45}, (1977)670.
\bibitem{8} M. Gronau, O. Hernandez, D. London and J. Rosner, Preprint,
TECHNION-PH-94-8, UdeM-LPN-TH-94-193, EFI-94-12; J. Silva and L. Wolfenstein,
Phys. Rev. {\bf D49}, (1994) 1151.
\bibitem{1} I.I. Bigi and A.I. Sanda, Nucl. Phys. {\bf B193}, (1981)85; {\bf
281}, (1987) 41;
For a review see: $B\; Decays$, edited by S. Stone, World
Scientific, 1994.
\bibitem{2} M. Gronau and D. London, Phys. Rev. Lett. {\bf 65}, (1990) 3381;
H. Lipkin, Y. Nir, H. Quinn and A. Snyder, Phys. Rev. {\bf D44}, (1991)1454.
\bibitem{3} N. G. Deshpande and Xiao-Gang He, Preprint, OITS-553, (Phys. Rev.
Lett. in press);  A. Deandrea, N. Di Bartilomeo, R. Gatto, F. Feruglio and J.
Nardulli, Phys. lett. {\bf B320}, (1994)170;
T. Hayashi, M. Matsuda and M. Tanimoto, Phys. Lett. {\bf B323}, (1994) 78;
D. Du and Z.Z. Xing, Phys. Lett. {\bf B280}, (1992)292; M. Gronau, Phys. Rev.
Lett. {\bf 63}, (1989)1451; B. Grinstein, Phys. lett. {\bf 229}, (1989)280;
D. London and R. Peccei, Phys. Lett. {\bf B223}, (1989)257; L.L. Chau and H.Y
Cheng, Phys. Rev. Lett. {\bf 59}, (1987)958; M. Gavela et al., Phys. Lett. {\bf
B154}, (1985)425.

\end{references}
\end{document}